# Resonant Critical Coupling of Surface Lattice Resonances with Fluorescent Absorptive Thin Film

Joshua T. Y. Tse[a*], Shunsuke Murai[a], and Katsuhisa Tanaka[a]

[a]Department of Material Chemistry, Graduate School of Engineering, Kyoto University, Katsura, Nishikyo-ku, Kyoto 615-8510, Japan

*Email: tse@dipole7.kuic.kyoto-u.ac.jp

**ABSTRACT**

Surface lattice resonance supported on nanoparticle arrays is a promising candidate in enhancing fluorescent effects in both absorption and emission. The optical enhancement provided by surface lattice resonance is primarily through the light confinement beyond the diffraction limit, where the nanoparticle arrays can enhance light-matter interaction for increased absorption as well as providing more local density of states for enhanced spontaneous emission. In this work, we optimize the in-coupling efficiency to the fluorescent molecules by finding the conditions to maximize the absorption, also known as the critical coupling condition. We studied the transmission characteristics and the fluorescent emission of a $TiO_2$ nanoparticle array embedded in an index-matching layer with fluorescent dye at various concentrations. A modified coupled-mode theory that describes the nanoparticle array was then derived and verified by numerical simulations. With the analytical model, we analyzed the experimental measurements and discovered the condition to critically couple light into the fluorescent dye, which is demonstrated as the strongest emission. This study presents a useful guide for designing efficient energy transfer from excitation beam to the emitters, which maximizes the external conversion efficiency.

- **INTRODUCTION**

Surface lattice resonance (SLR) is a photonic resonance supported on nanoparticles arranged in a one-, two- or three-dimensional periodic structure. [1-13] SLR have gained extensive attention through its potential in significantly improving light-matter interaction by introducing a confining the EM field beyond the diffraction limit and enhancing the field strength. [14-17] The nanoparticle arrays also possess great versatility in design and tunability, including the geometry of the array as well as the choice of materials used in the nanoparticle and the environment, and have found a wide range of applications in various fields, such as biosensing, [18,19] photovoltaic, [20,21] quantum optics, [22,23] nanolasing, [24,25] surface-enhanced spectroscopy, [26] etc.

The main theoretical model proposed to understand SLR have been the coupled dipole model, which approximates the nanoparticles as discrete dipoles and use the retarded dipole sum to calculate the collective behavior of the nanoparticle array. [27-32] An alternative to the coupled dipole model is the coupled mode theory (CMT), which studies the coupling of energy between two or more modes and ports, and have shown promising results in describing SLR as the coupling between the intrinsic mode in the nanoparticles and the diffraction mode presented by the periodic array geometry. [33-36] A particularly interesting part of CMT is the derivation towards the critical coupling condition. [37-39] Briefly, critical coupling is achieved by creating destructive interference between the non-resonant direct scattering and the resonant radiative decay at each outgoing ports to minimized the energy emitted outwards. In conventional CMT, this is achieved by two conditions: (1) the absorption and radiative decay rates should be the same, and (2) the magnitude and phase of the incidents at each port should be tuned properly. Achieving critical coupling in a steady-state allows the absorption to be maximized under energy conservation while the energy loss to emission is minimized.

Previous works have shown that SLR is capable in significantly enhancing the light-matter interaction of fluorescent dye molecules and can enhance both the in-coupling efficiency as well as directional out-coupling enhancement of the fluorescent emission. [40-44] The enhancement in in-coupling efficiency is due to the confinement of EM field into the vicinity of the nanoparticle array, and hence the localized and enhanced field strength would induce a stronger absorption by the fluorescent molecules. [40,45-47] On the other hand, the angle-dependent fluorescence enhancement is enabled by first enhancing the emission rate of the fluorescent molecules through the increased local density of state near the nanoparticle array, also known as the Purcell effect, and then the enhanced emission is radiatively out-coupled to specific angles through the diffraction with the nanoparticle array. [48-50]

In this work, we focus on using optimizing the in-coupling efficiency to fluorescent molecules mediated by the SLR. While the wavelength of the SLR should obviously match the absorption peak of the fluorescent molecule, and thus the periodicity as well as the nanoparticle's dimensions can be determined straightforwardly, what concentration of the fluorescent molecules best optimizes the energy transfer towards the fluorescent molecules is still unknown. Therefore, we approach this optimization problem based on the formalism of the temporal coupled-mode theory. We first examine the transmission characteristics of the SLR array when embedded in a fluorescent dye as well as the emission of the array, with different dye concentrations. We then derive a modified CMT that accurately describes the absorption behavior of the embedded nanoparticle array and compare the model with finite-difference time-domain (FDTD) simulated results. Lastly, we proceed to analyze the decay characteristics of the SLR mode and compare such with the emission features to evaluate the optimal point of coupling energy into the fluorescent molecules.

- **METHODS**

**Sample Preparation.**

We prepared the $TiO_2$ nanoparticle array on a $SiO_2$ glass substrate by electron-beam lithography followed by reactive ion etching. First, a $TiO_2$ thin layer with thickness of 90 nm was deposited on the silica glass substrate by RF (radio frequency) magnetron sputtering. The X-ray diffraction (XRD) pattern of the $TiO_2$ layer is shown in Supporting Information, Figure S1, in which the absence of distinct peaks indicates that the present TiO$_2$ layer is amorphous. Then, a resist (ZEP520A) was spin-coated and the nanohole array pattern was written by electron-beam lithography. After that, a Cr layer (120 nm) was deposited by electron-beam deposition, and the following lift-off process resulted in a Cr dot array pattern on the TiO$_2$ layer. Then, the TiO$_2$ layer was etched away by reactive ion etching with CHF$_3$ gas to make the nanoparticle array. The Cr dot mask was then removed by wet etching. The TiO$_2$ cylindrical nanoparticles have height $H$ = 90 nm and diameter $D$ = 130 nm, and are placed in a square lattice with periodicity $P$ = 380 nm. The SEM image of the fabricated TiO$_2$ nanoparticle array is shown in Figure 1a. Finally, a PMMA index-matching layer was spin-coated on the nanoparticle array. The thickness of the PMMA index-matching layer was measured to be $t$ = 460 nm ± 8 nm. The PMMA layer also contains a Lumogen dye (Lumogen F Red 305) of concentration $\rho$ = 0, 1, 2, 2.5, 3, 3.5, 4, 5, and 7 wt% (weight percent) ± 0.05 wt% respectively. ($\rho$ = 1 wt% corresponds to 11.23 mol/m$^3$, i.e., 0.0676 molecules/nm$^3$, and so on.) The Lumogen dye was selected because of its high photostability, which was of particular importance for quantitative comparison of emission intensity. Also, as will be shown later in Fig. 6d, the Lumogen dye shows no severe concentration quenching up to 7 wt%. Figure 1b illustrates the design of the sample. The geometry of the TiO$_2$ nanoparticle array was chosen such that the resonant wavelength of the SLR under normal incident would match the main absorption peak of the Lumogen dye near 580 nm (see Supporting Information, Figure S2).

**Transmissivity Measurements.**

We measured the polarization- and angle-resolved transmissivity bandstructure of the sample with the setup illustrated in Figure 1c. The prepared sample was mounted onto a rotation stage to measure the polarization- and angle-resolved transmissivity spectrum. The light from the stabilized tungsten-halogen lamp (Thorlabs SLS201L/M) was collimated with a 5X microscope objective. After that, the polarization of the incident light was controlled by a linear polarizer. The incident light then illuminated the sample from the PMMA side. Finally, the transmitted light was focused into an optical fiber that was connected to a spectrometer (Ocean Insight Flame-S, Grating Number 2, 25 μm slit). The transmission intensity was normalized against the intensity of the incident light to obtain the transmissivity.

**Emission Measurements.**

We measured the emission from the Lumogen dye excited at the resonant wavelength of the SLR under normal excitation, as illustrated in Figure 1d. The prepared sample was mounted on a goniometer for measurement of the fluorescent emission. The supercontinuum white light laser (Fianium WhiteLase micro) from an optical fiber was collimated by a collimator. The wavelength of the incident light was then controlled with a bandpass filter centered at 580 nm and width of 10 nm (Thorlabs FBH580-10). The incident light illuminated the sample from the substrate side at normal incident to excite the SLR and fluorescent absorption. The angle-resolved emission spectrum was then captured along the Γ-X direction with a lens and an optical fiber mounted on a moving arm, connecting to a spectrometer (Ocean Insight Flame-S, Grating Number 2, 25 μm slit).

**FDTD Simulation.**

We analyzed FDTD simulations on nanoparticle arrays with the same design as illustrated in Figure 1b. In the simulations, the nanocylinders have refractive index $n = 2.7$, diameter $D = 130$ nm and height $H = 100$ nm. The nanoparticles were placed in a square lattice with $P = 380$ nm and the $x$-, $y$-boundaries were periodic to simulate an infinite square lattice. The substrate is a semi-infinite layer with $n = 1.46$. The index-matching layer has thickness $t = 280$ nm and complex refractive index $n = 1.46 + i\kappa$. The imaginary part (extinction coefficient) of the complex refractive index $\kappa$ was varied to simulate the absorption effects over a range of different dye concentrations in the index-matching layer. A layer of vacuum ($n = 1$) was placed above the index-matching layer. The substrate and the vacuum extend through the perfectly-matching layers at the $z$-boundaries to simulate as semi-infinite layers.

- **RESULTS**

**Transmission Bandstructure Measurement.**

Figure 2 shows the TE- and TM-transmissivity bandstructure of the nanoparticle array with $\rho = 2$ wt% dye in the PMMA layer, measured along the Γ-X direction ($\phi = 0°$) from incident polar angle $\theta = -60°$ to $60°$. The SLR modes can be identified alongside the Rayleigh anomalies (RAs), which are the in-plane diffraction grating orders that can be calculated for the square lattice by:

$$\left(\frac{2\pi}{\lambda}\right)^2 = \left(\frac{2\pi}{\lambda}\sin\theta\cos\phi + n\frac{2\pi}{P}\right)^2 + \left(\frac{2\pi}{\lambda}\sin\theta\sin\phi + m\frac{2\pi}{P}\right)^2 \qquad (1)$$

where $\lambda$ is the wavelength, $\phi$ is the azimuthal angle and $(n, m)$ is the mode order of the RA [36]. In the TE-transmissivity bandstructure, only the $(1, 0)$ and $(-1, 0)$ SLR modes are excited. While in the TM-transmissivity bandstructure, the $(0, \pm 1)$ SLR modes are excited and the $(1, \pm 1)$ and $(-1, \pm 1)$ modes are also observed in larger incident angles. In the bandstructure, we observe the non-resonant part of the transmissivity being generally lower for wavelengths shorter than 600 nm, which is consistent with the absorption wavelength of the Lumogen dye (see Supporting Information, Figure S2). The measured linewidth of the $(1, 0)$ SLR mode also show an observable broadening as the SLR moves from longer wavelength to shorter wavelength, indicating a decrease in lifetime of the SLR. On the other hand, the $(0, \pm 1)$ SLR did not show significant change in the linewidth as it remained within the absorption window of the Lumogen dye. Figures 3 and 4 show the TE- and TM-bandstructure with $\rho = 1, 3, 5,$ and 7 wt%. The non-resonant absorption in the wavelengths shorter than 600 nm is observed to gradually increase with $\rho$. The corresponding linewidth of the SLR within the absorption window also display a gradual broadening as $\rho$ increase. These observed features suggest that the absorption of the SLR is dominantly controlled by the concentration of the Lumogen dye embedded in the PMMA index-matching layer and absorption can be optimized by tuning the dye concentration.

**Fluorescent Emission Measurements.**

In order to evaluate the absorption of the SLR mode, we compare the emission from the Lumogen dye between different $\rho$. Figure 5 shows the measured emission mapping of the SLR with $\rho = 0, 1, 3,$ 3.5, 5, and 7 wt%. In the emission spectra with $\rho$ = 1 to 7 wt%, we can see the broadband emission of the Lumogen dye covering approximately 570 to 700 nm. We also observed the out-coupling enhancement by the nanoparticle array along the (−1, 0) SLR mode, that appears as a sharp and dispersive line starting from $\lambda$ = 650 nm at $\theta$ = 10° and red-shifts as $\theta$ increases. The emission out-coupling by SLR [40, 41] and separated evaluation of in- and out-coupling had been discussed in previous publications. [51] A horizontal band of "emission" was also detected at 580 nm, which is the exact wavelength of the incident beam and the excited SLR. This horizontal band can be attributed to the scattering of the SLR mode from the inherent roughness in the nanoparticle array, since it is also observed in the sample without any fluorescent dye. The general trend of emission intensity with respect to the angle of emission follows the Lambert's cosine law describing ideal diffuse radiators.

We compare the emission strength of the SLR with different $\rho$ value at $\theta$ = 45°, at which the (−1, 0) diffraction order red-shifts to be out of emission range of the Lumogen dye. The emission spectra in Figure 6 are normalized to the measured intensity of the incident light through the 580 nm bandpass filter. Due to the increase in attenuation for higher dye concentration, the scattering peak at 580 nm decreased in magnitude as the dye concentration increase. The normalized emissions are shown in Figure 6a and the emission peak values near 610 nm are extracted and plotted as a function of $\rho$ in Figure 6b. We also measured the normalized emission spectra of the PMMA layer on the unstructured substrate as shown in Figure 6c. Because of the absence of nanostructure, the scattering of incident light is less than that for the array sample and thus no peak at $\lambda$ = 580 nm appears. The corresponding extracted emission peak values are plotted in Figure 6d. For the PMMA layer on the unstructured substrate, we can see that the fluorescent emission increases linearly with $\rho$ up until 4 wt%, which indicates the fluorescence efficiency of the dye is not affected by adding more dye. However, as $\rho$ is increased to 5 wt% or more, the fluorescent emission intensity drops below the linear trend, indicating a reduction in fluorescence efficiency per unit of dye. On the other hand, for the nanoparticle array, the emission follows a concave downwards trend and the emission intensity only increases until $\rho$ =

3.5 wt%, at which it is enhanced by 3.47 times compared to that from the same layer but on the unstructured substrate. Interestingly, further increasing $\rho$ to 4 wt% or more decreases the emission instead.

**Modified Coupled Mode Theory.**

In order to get a better understanding of the SLR, we model the behavior of the SLR on the nanoparticle array with a modified CMT. In a lossless system, the mode amplitude $a$ of the SLR mode can be described by the following equation: [35,52] $\frac{\text{d}}{\text{d}t}a = \left(i\omega_0 - \frac{\Gamma}{2}\right)a + \sqrt{\frac{\Gamma}{2}}\langle\kappa^*|s_+\rangle$, and the outgoing wave $|s_-\rangle = (s_{-,\text{TE},\uparrow} \quad s_{-,\text{TM},\uparrow} \quad s_{-,\text{TE},\downarrow} \quad s_{-,\text{TM},\downarrow})^\text{T}$ is described by $|s_-\rangle = \mathbf{C}|s_+\rangle + a\sqrt{\frac{\Gamma}{2}}|\kappa\rangle$, where $\omega_0$ is the resonant frequency and $\Gamma$ is the decay rate of the SLR. $|s_+\rangle = (s_{+,\text{TE},\uparrow} \quad s_{+,\text{TM},\uparrow} \quad s_{+,\text{TE},\downarrow} \quad s_{+,\text{TM},\downarrow})^\text{T}$ is the incident wave vector, the direct scattering matrix is

$$\mathbf{C} = \begin{pmatrix} t_{0,\text{TE}} & 0 & r_{0,\text{TE}} & 0 \\ 0 & t_{0,\text{TM}} & 0 & r_{0,\text{TM}} \\ r_{0,\text{TE}} & 0 & t_{0,\text{TE}} & 0 \\ 0 & r_{0,\text{TM}} & 0 & t_{0,\text{TM}} \end{pmatrix}$$

where $t_{0,\text{TE/TM}}$ and $r_{0,\text{TE/TM}}$ are the non-resonant transmissivity and reflectivity respectively, and $|\kappa\rangle = (\kappa_{\text{TE},\uparrow} \quad \kappa_{\text{TM},\uparrow} \quad \kappa_{\text{TE},\downarrow} \quad \kappa_{\text{TM},\downarrow})^\text{T}$ are the in-coupling constants. The subscripts TE/TM denotes the polarization and ↑/↓ indicates the upwards/downwards propagation direction of each port. To account for the influence on in/out-coupling and scattering from the absorption in the index-matching layer, we generalize the equations describing the SLR mode and the outgoing wave in the following form:

$$\frac{\text{d}}{\text{d}t}a = \left(i\omega_0 - \frac{\Gamma_\text{tot}}{2}\right)a + \sqrt{\frac{\Gamma_\text{rad}}{2}}\alpha\langle\kappa^*|s_+\rangle \text{ and} \tag{2}$$

$$|s_-\rangle = \beta\mathbf{C}|s_+\rangle + a\sqrt{\frac{\Gamma_\text{rad}}{2}}\alpha|\kappa\rangle \tag{3}$$

where $\alpha$ and $\beta$ are introduced to modify the in-coupling constants and the direct scattering matrix respectively. Note that $\alpha$ and $\beta$ should return to 1 when the index-matching layer is not lossy, which ensures that this modified CMT is consistent with the well-established lossless CMT. The $\Gamma$ are replaced by the total decay rate $\Gamma_{tot}$ and the radiative decay rate $\Gamma_{rad}$ correspondingly, where the total decay rate $\Gamma_{tot}$ is the sum of the radiative decay rate $\Gamma_{rad}$ and the absorptive decay rate $\Gamma_{abs}$. With the conditions of conservation of energy and time-reversal symmetry (see Supporting Information), we find that $\alpha = \sqrt[4]{1-A_0}$ and $\beta = \sqrt{1-A_0}$ where $A_0$ is the non-resonant absorptivity when the SLR mode is not excited. Therefore, we obtain the following equations that describe our nanoparticle array embedded in an absorptive thin film:

$$\frac{d}{dt}a(t) = \left(i\omega_0 - \frac{\Gamma_{rad}+\Gamma_{abs}}{2}\right)a(t) + \sqrt{\frac{\Gamma_{rad}}{2}}\sqrt[4]{1-A_0}\langle\kappa^*|s_+\rangle \text{ and} \quad (4)$$

$$|s_-\rangle = \sqrt{1-A_0}\,\mathbf{C}|s_+\rangle + a(t)\sqrt{\frac{\Gamma_{rad}}{2}}\sqrt[4]{1-A_0}|\kappa\rangle. \quad (5)$$

The TE/TM-polarized transmissivity and reflectivity derived from eqs 4 and 5 are:

$$T_{TE/TM} = (1-A_0)\left|t_{0,TE/TM} + \frac{\Gamma_{rad}}{2}\frac{\kappa^2_{TE/TM}}{i(\omega-\omega_0)+(\Gamma_{rad}+\Gamma_{abs})/2}\right|^2 \text{ and} \quad (6)$$

$$R_{TE/TM} = (1-A_0)\left|r_{0,TE/TM} + \frac{\Gamma_{rad}}{2}\frac{\kappa^2_{TE/TM}}{i(\omega-\omega_0)+(\Gamma_{rad}+\Gamma_{abs})/2}\right|^2, \quad (7)$$

under steady-state condition, which are Fano-like spectra. [36]

We then derive the absorptivity of the nanoparticle array by considering $A = 1 - \frac{\langle s_-|s_-\rangle}{\langle s_+|s_+\rangle}$, where we find the absorptivity to be (see Supporting Information):

$$A = A_0 + (1-A_0)\frac{\Gamma_{rad}\Gamma_{abs}}{2}\frac{|\langle\kappa^*|s_+\rangle|^2}{\langle s_+|s_+\rangle}\left|\frac{1}{i(\omega-\omega_0)+\frac{\Gamma_{rad}+\Gamma_{abs}}{2}}\right|^2, \quad (8)$$

under the steady-state solution, and:

$$A = A_0 + (1-A_0)\frac{2\Gamma_{rad}\Gamma_{abs}}{(\Gamma_{rad}+\Gamma_{abs})^2}\frac{|\langle\kappa^*|s_+\rangle|^2}{\langle s_+|s_+\rangle}, \quad (9)$$

at the resonant wavelength of the SLR mode.

From eq 9, we can see that there are two contributing factors for the absorption in the nanoparticle array, namely the non-resonant absorption and the resonant SLR absorption. While the contribution of the non-resonant absorption is straightforward, the SLR absorption is best pronounced when $\Gamma_{rad} = \Gamma_{abs}$, which optimizes the efficiency in coupling energy to the absorption. However, due to the $(1 - A_0)$ term and the correlation between $\Gamma_{abs}$ and $A_0$, critical coupling can only be achieved by balancing the effect of both factors. Since both $\Gamma_{abs}$ and $A_0$ are dependent on the dye concentration, it is helpful to look at different situations as the dye concentration increases. In the special case without the dye, since the $TiO_2$ nanoparticles also have a very low absorption, the absorption becomes negligible and the decay is primarily contributed by the radiative decay. As the "dye concentration", or in general the absorptivity of the thin film, increases but is still relatively low, $A_0$ is small and the absorption is mainly contributed by the absorption of the SLR mode. Further increasing the "dye concentration" would further increase $\Gamma_{abs}$ and $A_0$, which would result in an increasing absorption background. However, as can be seen in the modification factors $\alpha$ and $\beta$, the contribution of SLR absorption is impeded by the $(1 - A_0)$ factor. Also, as the $\Gamma_{abs}$ increases to larger than the $\Gamma_{rad}$, the efficiency in absorbing the energy coupled into the SLR mode also declines. The combined effect is the absorption would decrease and there exist an optimum point for the "dye concentration" to maximize absorption. Therefore, we want to increase the absorption from the SLR with increasing "dye concentration" to reach critical coupling but not to overshoot such that the non-resonant absorption impedes the coupling towards the SLR significantly.

**Numerical Simulation Results.**

Figure 7 shows the simulated transmissivity, reflectivity and absorptivity spectra at normal incident for selected $\kappa$ = 0, 0.01, 0.02. The transmissivity, reflectivity and absorptivity are fitted with eqs 6 to 8, respectively. The best fits are also plotted in Figure 7 as the black dashed lines. We can observe some features similar to the experimental results, namely the increase in non-resonant absorption and linewidth broadening as $\kappa$ increase. The total, radiative and absorptive decay rates obtained from the best fits are plotted against $\kappa$ in Figure 8a and the $A_0$ is also plotted against $\kappa$ in Figure 8b. From the fitted parameters, we can see that while the radiative decay rate $\Gamma_{rad}$ is constant against $\kappa$, the absorption decay rate $\Gamma_{abs}$ and the $A_0$ are both proportional to $\kappa$, thus resulting in the total decay rate $\Gamma_{tot}$ increasing linearly with $\kappa$. With the obtained parameters, we proceed to predict the absorptivity of the nanoparticle array with eq 9. The simulated absorptivity at the resonant wavelength of the SLR are plotted against $\kappa$ in Figure 8c, where the CMT model represented by the red solid line accurately predicted the absorptivity of the nanoparticle array. As shown in Figure 8c, the absorption first increases rapidly and reaches critical coupling at $\kappa = 0.011$, then it slowly rolls off as $\kappa$ further increases. At the peak, the absorptivity is enhanced by 7.96 times compared to the $A_0$, which approximates the absorptivity of the same layer on an unstructured substrate. This value is larger than that experimentally obtained emission intensity enhancement of 3.47 due to the difference in thickness between the fluorescent layer in experiment (460 nm) and that of simulations (280 nm). Since the nearfield of the SLR is concentrated near the nanoparticles, the dye further away from the nanoparticle contributes less in SLR absorption. However, as the non-resonant absorption contribution is uniform throughout the layer, the extra thickness contributes proportionally more towards the non-resonant absorption than SLR absorption and reduces the enhancement ratio. When we compare the absorption with the fitted decay rates in Figure 8a, in which we find the intersection between $\Gamma_{rad}$ and $\Gamma_{abs}$ at $\kappa = 0.0092$, we can see that although the contribution to absorption of the SLR is maximized at $\kappa = 0.0092$, critical coupling requires slightly increasing the $\kappa$ further from the intersection. This is because the non-resonant absorption still increases for increasing $\kappa$, and until the linear increase of the non-resonant absorption is overwhelmed by the roll off in absorption of the SLR, we can further increase the absorption.

To further confirm that our analytical model adequately describes SLR supported on this type of embedded nanoparticle array with an absorptive layer, we performed numerical simulations for nanoparticle arrays with a variety of geometrical parameters. Nanoparticle arrays with different nanoparticle diameter $D = 110, 150$ nm (labeled as D110, D150), square arrays with different periodicity $P = 300, 350$ nm (labeled as P300, P350), and hexagonal nanoparticle arrays ($D = 110$, 130 nm, $P$ = 380 nm, labeled as H110, H130) were analyzed with the simulations to show the versatility of our model. Similarly, the simulated transmissivity, reflectivity and absorptivity were fitted to obtain the total, radiative and absorptive decay rates, as well as the non-resonant absorption (see Supporting Information, Figure S3 and S4). The parameters are then used to predict the absorptivity of each configuration. The predicted absorptivity and the simulated absorptivity are plotted against $\kappa$ in Figures 8d–f to show the model accurately predicts the absorptivity. Interestingly, we can see in Figures 8d–f that only D110, P350 and H110 show critical coupling, while the other geometries show no absorption maximum but a continuous increase in absorption as $\kappa$ increases. This difference can be attributed to the value of $\kappa$ where $\Gamma_{\text{rad}}$ and $\Gamma_{\text{abs}}$ intersects. As shown in the Supplementary Information, in all cases where critical coupling can be reached, $\Gamma_{\text{rad}}$ and $\Gamma_{\text{abs}}$ intersects at a relatively small $\kappa$, which indicates that the absorption contribution of the SLR takes place at a small $\kappa$. Moreover, $\Gamma_{\text{rad}}$ and $\Gamma_{\text{abs}}$ intersecting at a small $\kappa$ also imply that the $\frac{2\Gamma_{\text{rad}}\Gamma_{\text{abs}}}{(\Gamma_{\text{rad}}+\Gamma_{\text{abs}})^2}$ factor decays relatively fast when compared to the gradual increase in $A_0$, as a function of $\kappa$. Since the rate of increase in $A_0$ is almost independent of the geometry of the array, the rate of decrease of the SLR absorption after the maximum dictates the existence of the absorption maximum observed in Figures 8d–f. Therefore, when the decay rates intersect at a large $\kappa$, the roll off in absorption of the SLR, given by the $\frac{2\Gamma_{\text{rad}}\Gamma_{\text{abs}}}{(\Gamma_{\text{rad}}+\Gamma_{\text{abs}})^2}$ and $(1 - A_0)$ factors, is too slow to compensate the increase in $A_0$ to show an absorption maximum. To conclude, we can see that our model is adequate in describing the decay mechanism and absorption of a variety of nanoparticle arrays embedded in an absorptive layer.

- **DISCUSSION**

We then use this CMT framework to analyze the experimentally measured spectrum. The best fit of the transmission spectrum of each sample are shown in Figure 9a as dashed lines. The decay rates obtained from the fittings are then plotted in Figure 9b for comparison. From Figure 9b, we can see that while the $\Gamma_{\text{rad}}$ is mostly constant over different dye concentrations, $\Gamma_{\text{abs}}$ increases proportionally with the dye concentration $\rho$, thus resulting in a linear increase of the total decay rate $\Gamma_{\text{tot}}$, and correspondingly the linewidth of the SLR against the dye concentration. When we compare these results with the FDTD simulated decay rates, we can see that $\Gamma_{abs}$ did not drop to zero when there is no dye at all. This can be attributed to the random scattering loss from irregularities of the nanoparticle array as the energy lost to random scattering would not be captured in the $\Gamma_{rad}$ but as part of $\Gamma_{abs}$. This energy loss to random scattering also contributes to the discrepancy in the enhancement ratio when comparing the experimentally obtained emission enhancement of 3.47 to the numerically simulated absorptivity enhancement of 7.96. As the random scattering provides an extra pathway for the SLR to decay and contributes to the total decay rate, more dye is needed to reach critical coupling when compared to an ideal case such as in simulations. This corresponds to a higher emission intensity on the unstructured substrate while the maximum SLR emission is constrained by the energy loss to random scattering, thus lowering the enhancement ratio. On the other hand, a possible explanation of the small drift in the $\Gamma_{rad}$ is that the change in contrast of the effective refractive index between the nanoparticle and the PMMA layer is affected by the change in dye concentration, as it may create a shift in the effective refractive index of the PMMA layer.

Since the emission strength is expected to positively correlate with the absorption to the Lumogen dye, we can infer the trends in absorption through analyzing the trends in emission intensity. First, if we consider the fluorescent emission of the PMMA layer on the unstructured substrate shown in Figure 6d, we can see that the emission is linear with $\rho$ up until $\rho$ = 4 wt%, which suggests that the absorption is also proportional to $\rho$. On the other hand, the emission from the SLR sample, as shown in Figure 6b, increases initially and is maximized at $\rho$ = 3.5 wt%, but further increasing $\rho$ to 4 wt% or more leads to a gradual decrease in emission. When we compare this with the fitted decay rates in Figure 9b, which shows an intersection between $\Gamma_{rad}$ and $\Gamma_{abs}$ occurring at $\rho$ = 2.9 wt%, indicating the position where the absorption contribution by the SLR is maximized. Based on what we learned from eq 9 and the simulation results, we know that while the intersection is at $\rho$ = 2.9 wt%, the position of the critical coupling condition is expected to be at slightly larger $\rho$. This suggests that there exists a maximum in absorption around $\rho$ = 3 to 4 wt%, which matches with our observation that the maximum emission is at $\rho$ = 3.5 wt%. Therefore, we can conclude that we derived the situation critically coupled into the SLR at $\rho$ = 3.5 wt% dye concentration.

By looking closely at the decay mechanism of the SLR, we can also evaluate the advantages and disadvantages of using non-absorptive $TiO_2$ nanoparticles over plasmonic metallic nanoparticles. By having effectively zero absorption in the nanoparticles, $TiO_2$ nanoparticles can direct all absorption towards the fluorescent molecules as that is the only absorbing mechanism in the system. On the other hand, for metallic nanoparticles which inherently absorb due to ohmic dissipation, the energy coupled into the SLR mode have three paths to decay, namely radiative decay, absorption by the metallic nanoparticle and absorption by the fluorescent dye. While the energy loss to the radiative decay can be managed and minimized by critical coupling, the absorption to the metal cannot be eliminated and reduces the overall efficiency in energy transfer towards the fluorescent emitters. Besides, the contrast in effective refractive index between nanoparticles and the embedded environment is also more pronounced with metallic nanoparticles than dielectric nanoparticles, which contributes to a larger radiative decay rate for the metallic nanoparticles with the same geometry. This discrepancy means that the absorption decay rate should also be tuned higher in order to match the higher radiative decay rate and achieve critical coupling. This leads to two results: a higher total decay rate at critical coupling and requiring a higher dye concentration to reach critical coupling. The higher total decay rate, also observable as a broader resonant linewidth, will aid in in-coupling more energy from a broadband excitation beam, but insignificant under monochromatic excitation. However, requiring dye concentration that is too high may also lead to no critical coupling, similar to some of the simulated results. On the other hand, we should also consider the difference in nearfield enhancement when evaluating the required dye concentration. SLR with metallic nanoparticles possess a significantly stronger EM field confinement, which corresponds to a much stronger field enhancement near the nanoparticles compared to dielectric nanoparticles, such as the $TiO_2$ nanoparticles used in this study. This stronger field enhancement gives a stronger light-matter interaction between the SLR nearfield and the fluorescent dye, leading to a higher absorption decay rate contribution with the same dye concentration compared to the dielectric counterpart. Therefore, the required dye concentration to achieve critical coupling on metallic nanoparticles depends on the two counteracting factors and the comparison with dielectric nanoparticles may differ depending on the geometry.

- **CONCLUSION**

In summary, we presented a rational scheme to optimize the in-coupling efficiency to the index-matching layer with fluorescent molecules on a $TiO_2$ nanoparticle array. First, we studied the transmissivity bandstructure of the SLR as well as the fluorescent emission from the array. When we compare the fluorescent emission between different dye concentrations, we discovered that the emission intensity did not increase linearly with the dye concentration, but it reached a maximum at 3.5 wt% with a 3.47 times enhancement by SLR, and the intensity decreased for higher concentration. To understand this phenomenon, we derived an analytical model based on the CMT formalism and verified that the model with FDTD simulation that it accurately describes the absorption behavior of the index-matching layer. Then, we used this analytical model to analyze the transmissivity spectra and the fluorescent emission, and found the decay rates behaves similar to the simulation results. Finally, we analyzed the trend of the fluorescent emission and find the critical coupling condition of the experimental sample to be at $\rho = 3.5$ wt%. This optimizes the energy transfer efficiency from the excitation beam to the emitters, which in turns aid in maximizing the external conversion efficiency.

- **ASSOCIATED CONTENT**

**Supporting Information**

The supporting Information is available free of charge at [link].

X-ray diffraction pattern of $TiO_2$ layer, absorption and emission spectrum of the Lumogen dye, derivations of $\alpha$, $\beta$ and the absorptivity $A$, and FDTD simulations of other geometries (PDF)

▪ **ACKNOWLEDGEMENTS**

The authors acknowledge financial support from Kakenhi (22H01776, 22K18884, 21H04619), MEXT, Japan, CASIO science promotion foundation, and IZUMI science and technology foundation.

Figure 1. (a) The top-down SEM image of the fabricated sample. The scale bar is 500 nm. (b) The illustration of the fabricated TiO$_2$ nanoparticle array. The TiO$_2$ nanoparticles are cylindrical with height *H* and diameter *D*, and are placed in a square lattice with periodicity *P*. The nanoparticle array is covered by a PMMA index-matching layer of thickness *t*. (c) The schematic of the measurement setup for transmissivity measurements. The light from the stabilized Tungsten-Halogen lamp is collimated with a 5X microscope objective (OBJ) and the polarization is controlled by a polarizer (P1). The light is then directed through the prepared sample (S) from the PMMA side and focused into an optical fiber connected to a spectrometer (SP) by a lens (L1). (d) The schematic of the measurement setup for fluorescent emission measurements. The supercontinuum white light laser is collimated by a collimator (C) and passes through a bandpass filter (BP). The light is then shone on the sample (S) from the substrate side at normal incident to excite the SLR and fluorescent absorption. The emission at *θ* is then focused by a lens (L2) into the optical fiber connected to the spectrometer (SP).

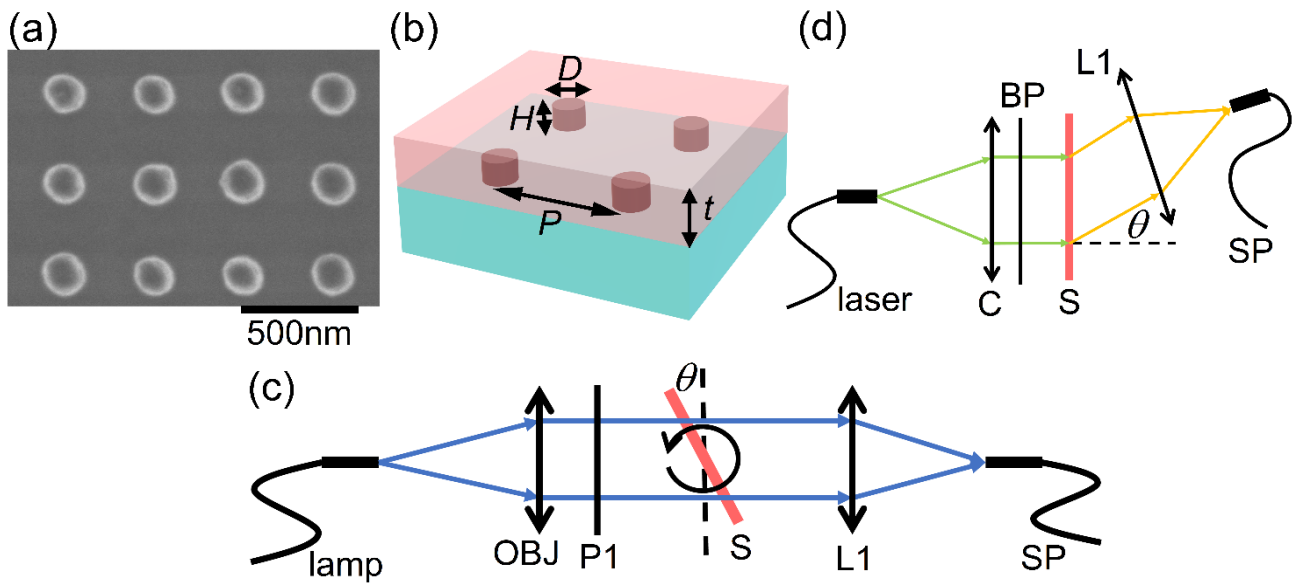

Figure 2. The (a) TE- and (b) TM-transmissivity bandstructure of the nanoparticle array with $\rho = 2$ wt%. The SLR modes (blue lines) are identified alongside the RAs (white dashed lines).

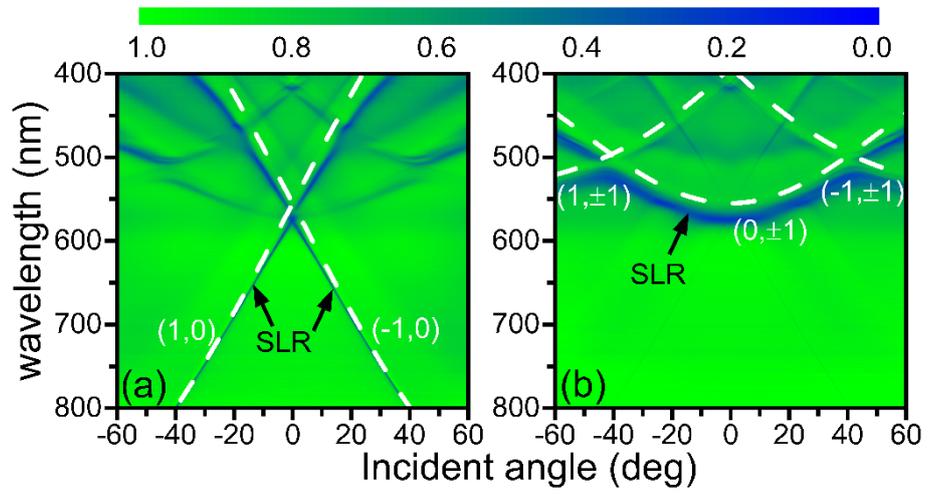

Figure 3. The TE-transmissivity bandstructure of the nanoparticle array with $\rho$ = (a) 1, (b) 3, (c) 5, (d) 7 wt%.

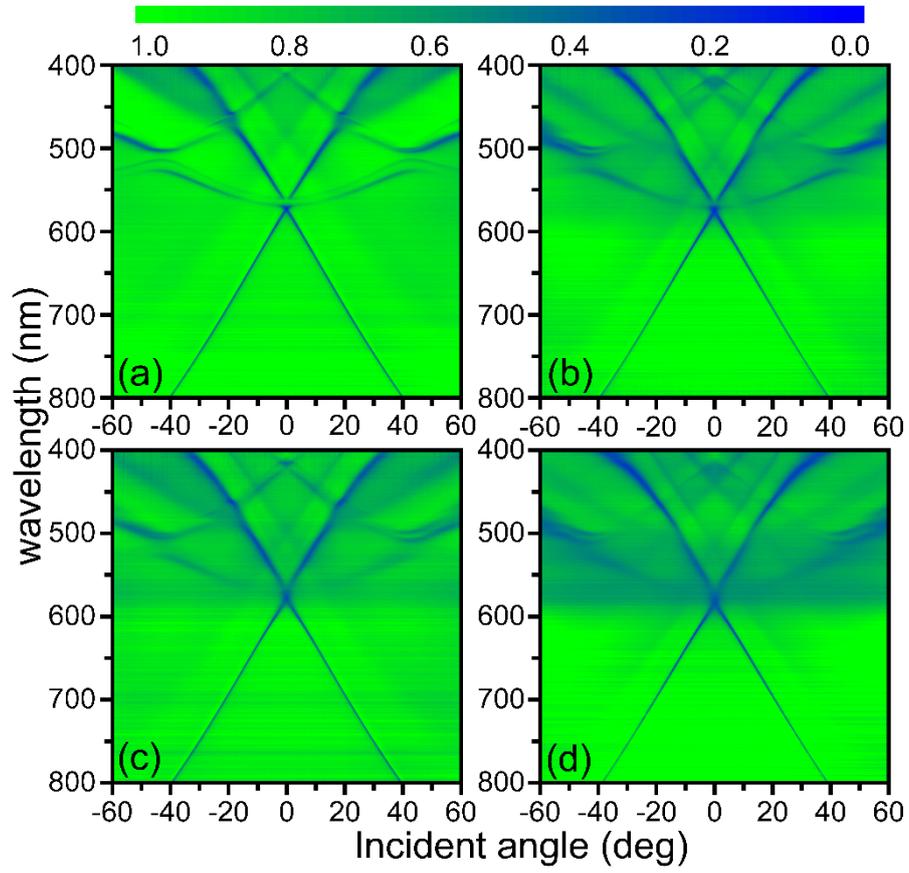

Figure 4. The TM-transmissivity bandstructure of the nanoparticle array with *ρ* = (a) 1, (b) 3, (c) 5, (d) 7 wt%.

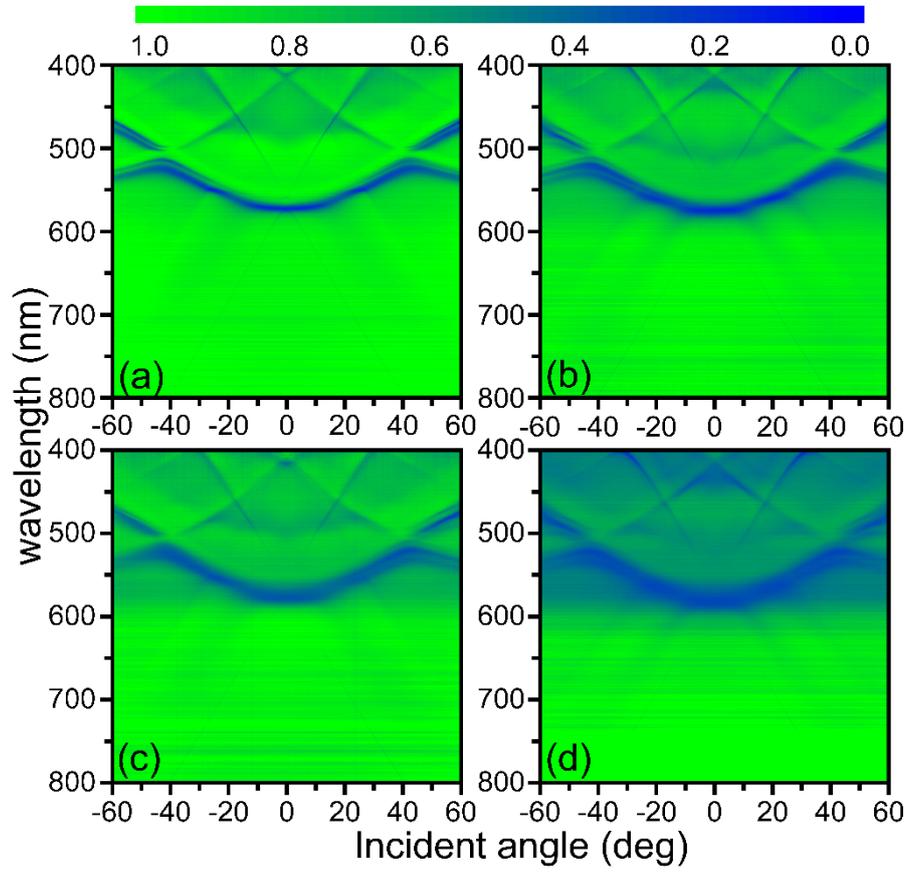

Figure 5. The emission mapping of the nanoparticle array with $\rho$ = (a) 0, (b) 1, (c) 3, (d) 3.5 (e) 5, (f) 7 wt%.

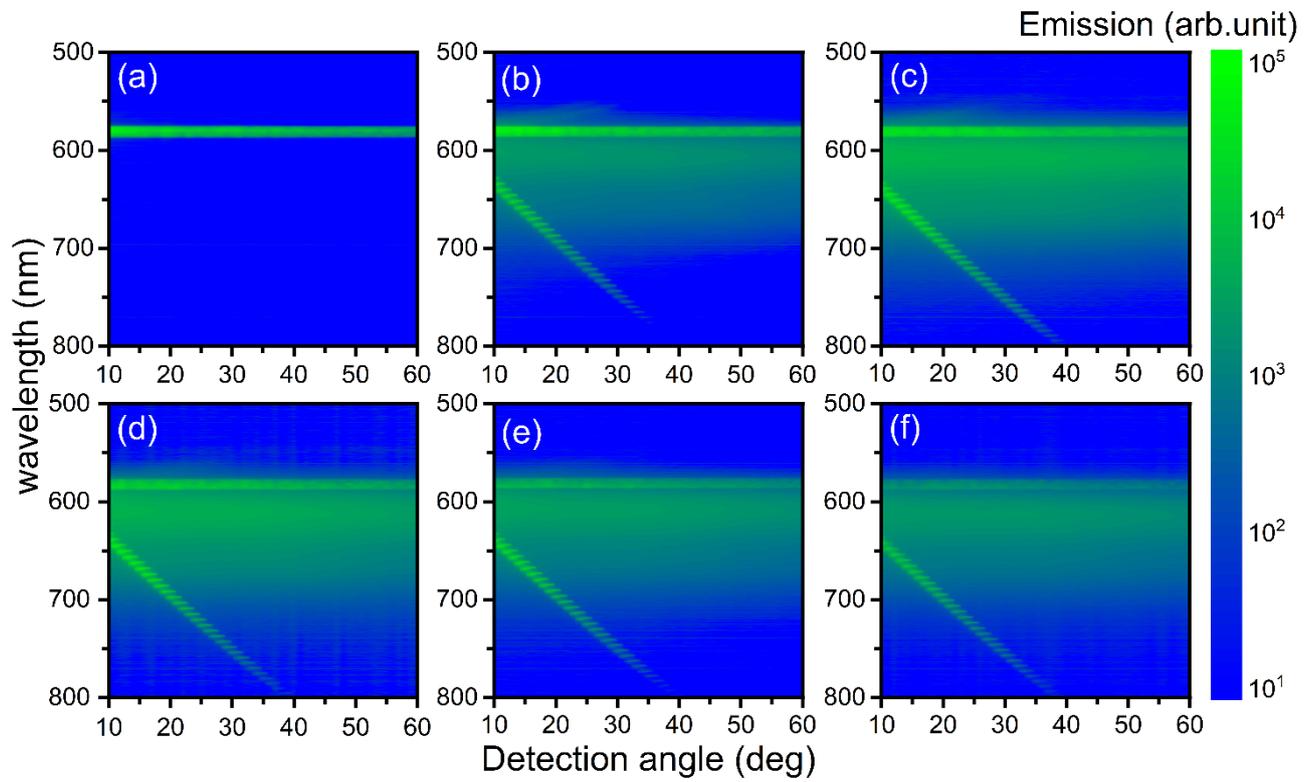

Figure 6. (a) The normalized emission spectra of the nanoparticle array measured at $\theta = 45°$, and (b) the corresponding emission peak values near 610 nm plotted as a function of $\rho$. (c) The normalized emission spectra of the PMMA layer on the unstructured substrate measured at $\theta = 45°$, and (d) the corresponding emission peak values plotted as a function of $\rho$. The dashed line indicates the linear trend at $\rho = 0 - 4$ wt%.

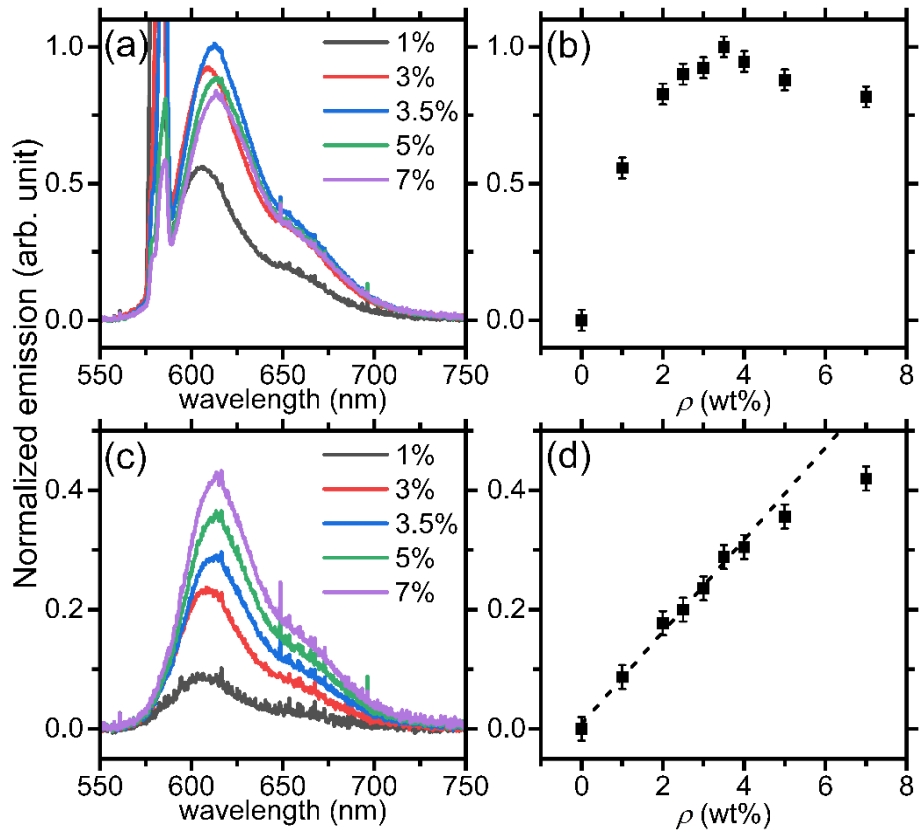

Figure 7. The FDTD simulated transmissivity, reflectivity and absorptivity spectra at normal incident for selected $\kappa$ = (a) 0, (b) 0.01, (c) 0.02. The best fit with the CMT are plotted as the dashed lines.

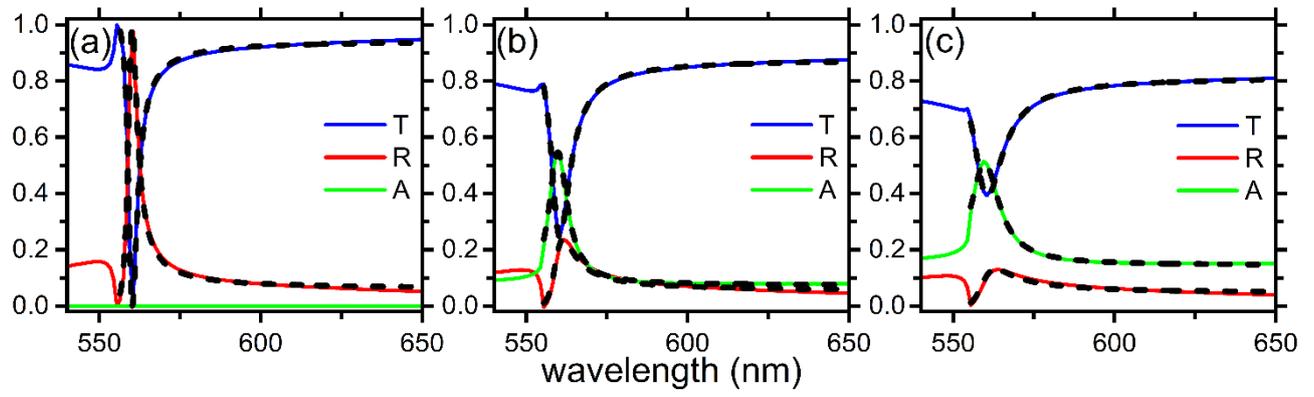

Figure 8. The (a) $\Gamma_{tot}$ (black squares), $\Gamma_{rad}$ (red circles), $\Gamma_{abs}$ (blue triangles) and (b) $A_0$ are plotted as a function of $\kappa$. The trends of the parameters are fitted by the straight lines as shown. (c) The simulated absorptivity at the resonant wavelength of the SLR are plotted as a function of $\kappa$. The red solid line is the CMT predicted absorptivity from the fittings. (d–f) The CMT predicted absorptivity and the FDTD simulated absorptivity of (d) D110, D150, (e) P300, P350, (f) H110 and H130 are plotted as a function of $\kappa$. The solid lines show the CMT prediction while the symbols show the FDTD simulated results.

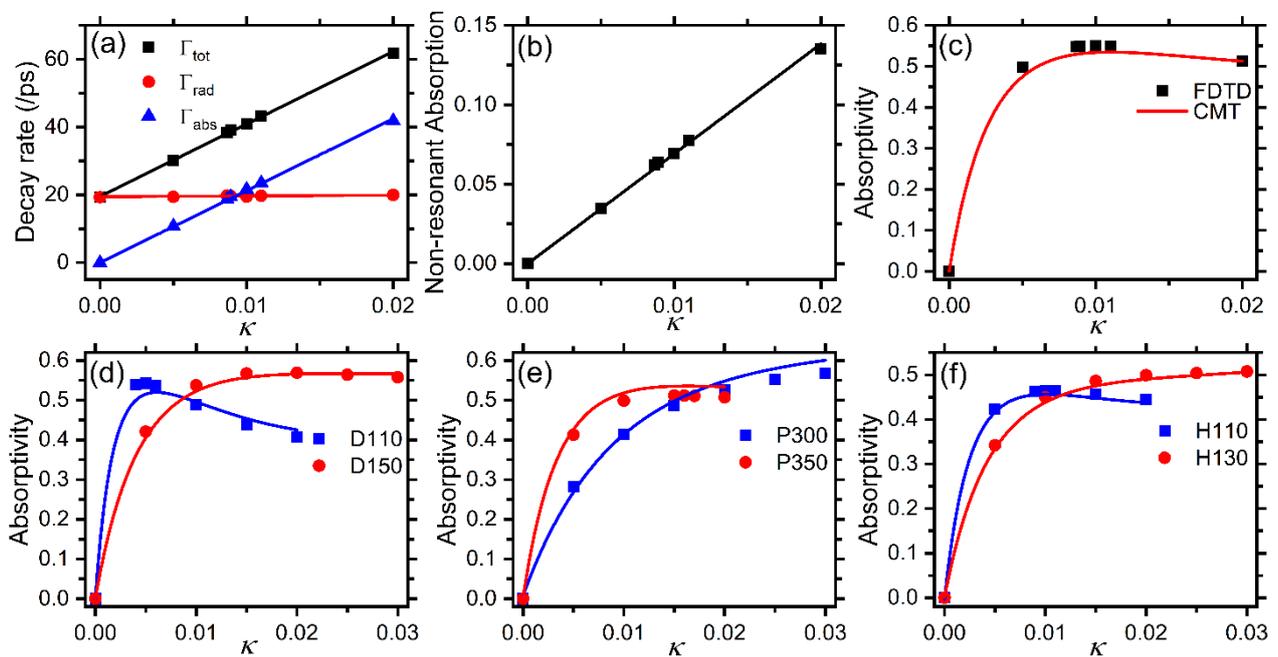

Figure 9. (a) The transmissivity spectra of the SLR of $\rho$ = 0, 1, 3, 5, and 7 wt% at normal incidence. The best fit of the spectra are plotted as dashed lines. (b) The $\Gamma_{tot}$ (black squares), $\Gamma_{rad}$ (red circles) and $\Gamma_{abs}$ (blue triangles) are plotted as a function of $\rho$. The symbols are obtained from the fitting curves in (a), and the solid lines are fits to the CMT model.

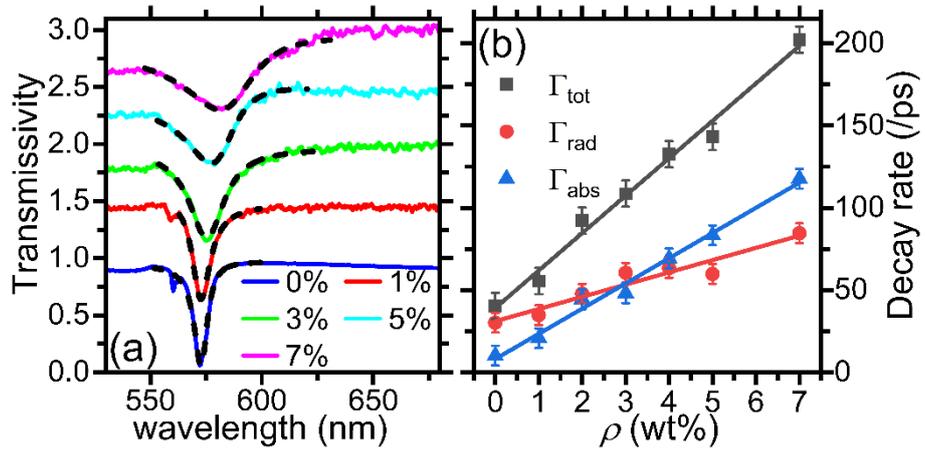

# Supporting Information

## Resonant Critical Coupling of Surface Lattice Resonances with Fluorescent Absorptive Thin Film

Joshua T. Y. Tse*, Shunsuke Murai, and Katsuhisa Tanaka

Department of Material Chemistry, Graduate School of Engineering, Kyoto University, Katsura, Nishikyo-ku, Kyoto 6158510, Japan

*Email: tse@dipole7.kuic.kyoto-u.ac.jp

### A. X-RAY DIFFRACTION PATTERN OF TiO$_2$ LAYER

The X-ray diffraction (XRD) pattern of the TiO$_2$ layer is measured after sputtering. As shown in Figure S1, the absence of distinct peaks indicates that the TiO$_2$ layer is amorphous.

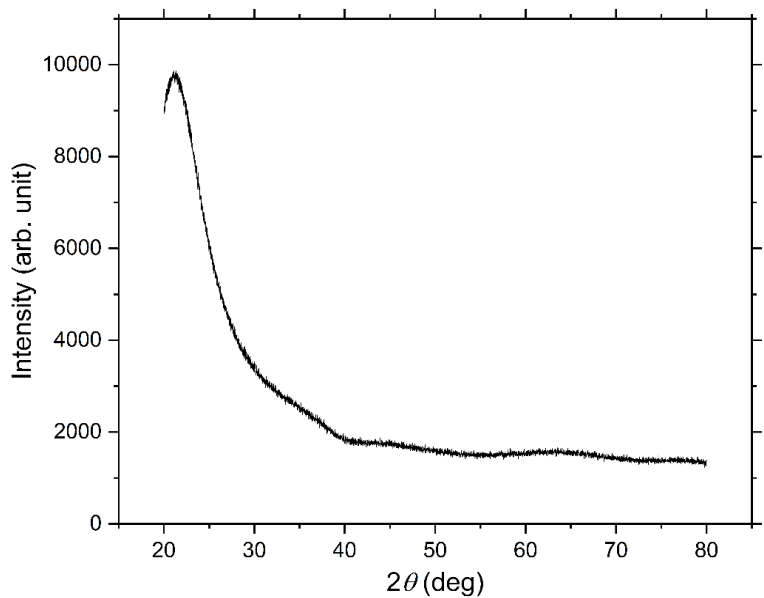

Figure S1. The XRD pattern of the as-deposited TiO$_2$ layer.

### B. ABSORPTION AND EMISSION SPECTRUM OF LUMOGEN DYE

The absorption and emission spectra of the Lumogen F Red 305 dye were measured as a reference. A PMMA layer of thickness $t = 460$ nm with embedded Lumogen dye of concentration $\rho = 5$ wt% was prepared on an unstructured SiO$_2$ glass substrate. Figure S2 shows the measured absorption and emission of the Lumogen dye normalized to the respective absorption and emission peak values. The transmissivity spectrum of the SLR at normal incident with $\rho = 0$ wt% is also plotted to show the SLR matches the wavelength of the absorption peak of the dye.

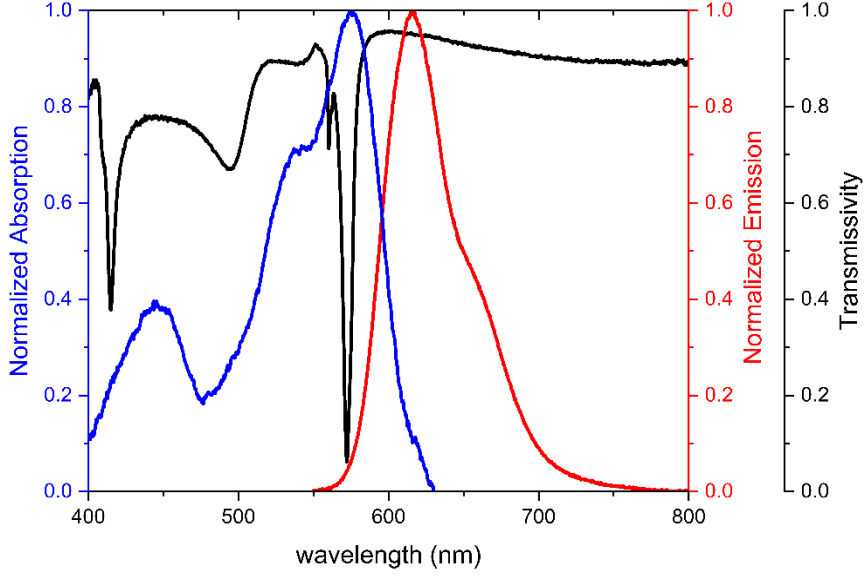

Figure S2. The normalized absorption (blue) and emission (red) spectra of the Lumogen dye. The transmissivity (black) of the SLR at normal incident with $\rho = 0$ wt% is also plotted for comparison.

## C. DERIVATIONS OF $\alpha$ AND $\beta$

We propose the following equations to describe the SLR with an absorptive index-matching layer:

$$\frac{d}{dt}a(t) = \left(i\omega_0 - \frac{\Gamma_{rad}+\Gamma_{abs}}{2}\right)a(t) + \sqrt{\frac{\Gamma_{rad}}{2}}\alpha\langle\kappa^*|s_+(t)\rangle \tag{S1}$$

$$|s_-(t)\rangle = \beta C|s_+(t)\rangle + a(t)\sqrt{\frac{\Gamma_{rad}}{2}}\alpha|\kappa\rangle \tag{S2}$$

where $\alpha$ and $\beta$ are introduced to modify the in-coupling constants and the direct scattering matrix respectively. The use of $\alpha$ and $\beta$ is limited to indicating the loss due to the absorption of the index-matching layer, and when the index-matching layer is not lossy, $\alpha$ and $\beta$ should return to 1. By considering the time-reversal symmetry of the Maxwell's Equations, in which the substitutions $\{\mathbf{E}(\mathbf{r},t), \mathbf{H}(\mathbf{r},t)\} \to \{\mathbf{E}(\mathbf{r},-t), -\mathbf{H}(\mathbf{r},-t)\}$ and $\{\tilde{\varepsilon},\tilde{\mu}\} \to \{\tilde{\varepsilon}^*,\tilde{\mu}^*\}$ would give another set of solution, the following transformations:

$$\{a(t), |s_\pm(t)\rangle\} \to \{a^*(-t), |s_\mp^*(-t)\rangle\} \text{ and } \{\Gamma_{abs}, \alpha, \beta\} \to \left\{-\Gamma_{abs}, \frac{1}{\alpha}, \frac{1}{\beta}\right\} \tag{S3}$$

should also give another valid solution in CMT. The sign of $\Gamma_{abs}$ is flipped to make the absorption decay an optical gain while $\alpha$ and $\beta$ are inverted because they are multiplicative factors. Therefore, we obtain another set of equations governing $a$ and $|s_\pm\rangle$:

$$\frac{d}{dt}a^*(-t) = \left(i\omega_0 - \frac{\Gamma_{rad}-\Gamma_{abs}}{2}\right)a^*(-t) + \sqrt{\frac{\Gamma_{rad}}{2}}\frac{1}{\alpha}\langle\kappa^*|s_-^*(-t)\rangle \qquad (S4)$$

$$|s_+^*(-t)\rangle = \frac{1}{\beta}\mathbf{C}|s_-^*(-t)\rangle + a^*(-t)\sqrt{\frac{\Gamma_{rad}}{2}}\frac{1}{\alpha}|\kappa\rangle \qquad (S5)$$

Here, we consider the case where the system is initially excited at $t = 0$ and has no external incident light, that is $a(t = 0) \neq 0$ and $|s_+(t)\rangle = 0$, we get:

$$\frac{d}{dt}a(t) = \left(i\omega_0 - \frac{\Gamma_{rad}+\Gamma_{abs}}{2}\right)a(t) \qquad (S6)$$

$$|s_-(t)\rangle = a(t)\sqrt{\frac{\Gamma_{rad}}{2}}\alpha|\kappa\rangle \qquad (S7)$$

$$-\frac{d}{dt}a^*(t) = \left(i\omega_0 - \frac{\Gamma_{rad}-\Gamma_{abs}}{2}\right)a^*(t) + \sqrt{\frac{\Gamma_{rad}}{2}}\frac{1}{\alpha}\langle\kappa^*|s_-^*(t)\rangle \qquad (S8)$$

$$0 = \frac{1}{\beta}\mathbf{C}|s_-^*(t)\rangle + a^*(t)\sqrt{\frac{\Gamma_{rad}}{2}}\frac{1}{\alpha}|\kappa\rangle \qquad (S9)$$

Substituting Eq. S7 into Eq. S9, we find:

$$\frac{1}{\beta}\mathbf{C}|\kappa^*\rangle = -\frac{1}{\alpha^2}|\kappa\rangle \qquad (S10)$$

By calculating the sum of Eq. S6 and Eq. S8's complex conjugate, and then substituting Eq. S7, we find that $\langle\kappa|\kappa\rangle = 2$, which is the same condition as in the lossless case. Assuming that the direct scattering matrix remains as a unitary matrix to be consistent with the lossless case, $\mathbf{C}^\dagger\mathbf{C} = \mathbf{I}$, we evaluate the magnitude of Eq. S10 by computing its inner product with itself and find that:

$$\alpha^2 = \beta. \qquad (S11)$$

We then consider the energy conservation of the system with the following equation:

$$\frac{d}{dt}|a|^2 = \langle s_+|s_+\rangle - \langle s_-|s_-\rangle - \Gamma_{abs}|a|^2 - \text{Abs}_{non-res} \qquad (S12)$$

where we require that the change in optical energy in the SLR resonance equals the incident power subtracted by the outgoing power, the optical absorption of the SLR mode and the non-resonance absorbing loss $\text{Abs}_{non-res}$. We consider the case where the SLR mode is not excited, that is $a(t) = 0$, by substituting Eq. S2, we find that $\langle s_+|s_+\rangle = \beta^2\langle s_+|\mathbf{C}^\dagger\mathbf{C}|s_+\rangle + \text{Abs}_{non-res}$. Since we define the non-resonant absorptivity as $A_0 = \frac{\text{Abs}_{non-res}}{\langle s_+|s_+\rangle}$ when the SLR mode is not excited, we derive that $A_0 = 1 - \frac{\beta^2\langle s_+|\mathbf{C}^\dagger\mathbf{C}|s_+\rangle}{\langle s_+|s_+\rangle} = 1 - \beta^2$, or $\beta = \sqrt{1-A_0}$ and $\alpha = \sqrt[4]{1-A_0}$.

## D. DERIVATION OF ABSORPTIVITY $A$

We compute the absorptivity by considering $A = 1 - \frac{\langle s_-|s_-\rangle}{\langle s_+|s_+\rangle}$, where by substituting Eq. 3 we find:

$$A = 1 - \frac{\left(\sqrt{1-A_0}\langle s_+|C^\dagger + a^*\sqrt{\frac{\Gamma_{rad}}{2}}\sqrt[4]{1-A_0}\langle\kappa|\right)\left(\sqrt{1-A_0}C|s_+\rangle + a\sqrt{\frac{\Gamma_{rad}}{2}}\sqrt[4]{1-A_0}|\kappa\rangle\right)}{\langle s_+|s_+\rangle}. \quad (S13)$$

By considering the steady-state solution of Eq. 2, we find that the mode amplitude is:

$$a = \frac{\sqrt{\frac{\Gamma_{rad}}{2}}\sqrt[4]{1-A_0}\langle\kappa^*|s_+\rangle}{i(\omega-\omega_0)+\frac{\Gamma_{rad}+\Gamma_{abs}}{2}}. \quad (S14)$$

We can then substitute Eq. S14 into Eq. S13, combining with equations derived in Section C, we simplify and obtain the equation:

$$A = A_0 + (1-A_0)\frac{\Gamma_{rad}\Gamma_{abs}}{2}\frac{|\langle\kappa^*|s_+\rangle|^2}{\langle s_+|s_+\rangle}\left|\frac{1}{i(\omega-\omega_0)+\frac{\Gamma_{rad}+\Gamma_{abs}}{2}}\right|^2. \quad (S15)$$

Therefore, at the resonant frequency, where $\omega = \omega_0$, we get:

$$A = A_0 + (1-A_0)\frac{2\Gamma_{rad}\Gamma_{abs}}{(\Gamma_{rad}+\Gamma_{abs})^2}\frac{|\langle\kappa^*|s_+\rangle|^2}{\langle s_+|s_+\rangle}. \quad (S16)$$

### E. FDTD SIMULATIONS OF OTHER GEOMETRIES

We further carried out numerical simulations for nanoparticle arrays with different geometrical parameters. The parameters used are summarized in Table S1 with the corresponding labels used to identify each configuration. All other simulation parameters are identical to the parameters as described in Methods. The fitted total, radiative and absorptive decay rates, and the non-resonant absorption are plotted as a function of $\kappa$ in Figure S3. The non-resonant absorptions are plotted as a function of $\kappa$ in Figure S4. The trends of the decay rates and the $A_0$ are similar to the primary configuration as described in the main text.

Table S1. The geometrical parameters used in simulations of different nanoparticle arrays.

| $D$ (nm) | $P$ (nm) | Lattice type | Label |
|---|---|---|---|
| 110 | 380 | square | D110 |
| 150 | 380 | square | D150 |
| 130 | 300 | square | P300 |
| 130 | 350 | square | P350 |
| 110 | 380 | hexagonal | H110 |
| 130 | 380 | hexagonal | H130 |

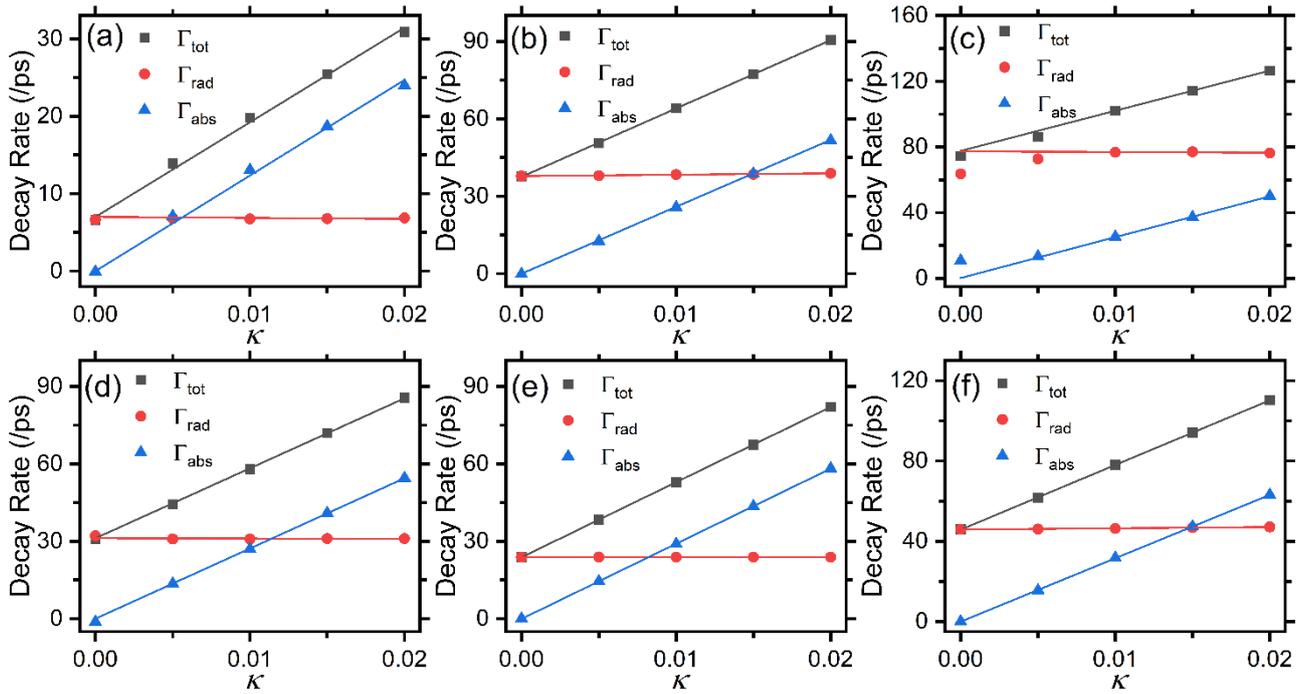

Figure S3. The fitted $\Gamma_{tot}$ (black squares), $\Gamma_{rad}$ (red circles), $\Gamma_{abs}$ (blue triangles) of configurations (a) D110, (b) D150, (c) P300, (d) P350, (e) H110, and (f) H130 are plotted as a function of $\kappa$. The trends of the parameters are fitted by the straight lines as shown.

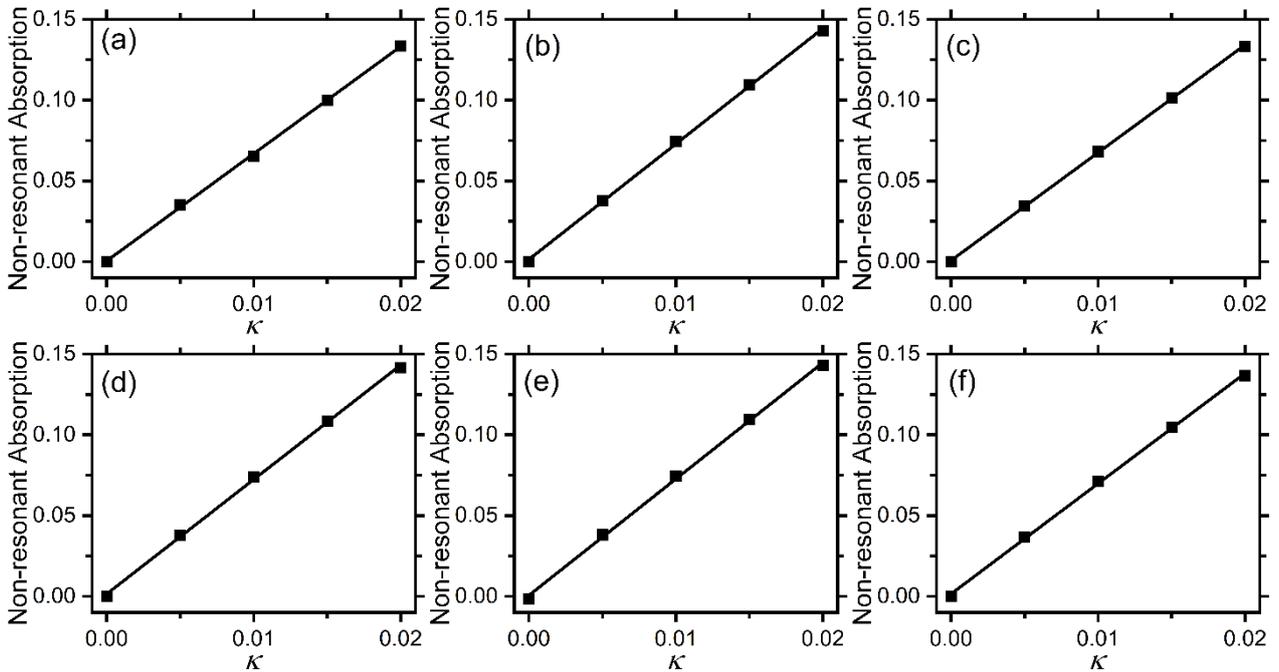

Figure S4. The fitted $A_0$ of configurations (a) D110, (b) D150, (c) P300, (d) P350, (e) H110, and (f) H130 are plotted as a function of $\kappa$. The trends of the parameters are fitted by the straight lines as shown.